\documentclass[onecolumn, twoside, notitlepage, superscriptaddress, nofootinbib, floatfix]{revtex4-1}

\usepackage[utf8]{inputenc} 
\usepackage[T1]{fontenc}    
\usepackage{hyperref}       
\usepackage{url}            
\usepackage{booktabs}       
\usepackage{amsfonts}       
\usepackage{nicefrac}       
\usepackage{microtype}      
\usepackage{graphicx} 
\usepackage{amsmath}
\usepackage{BOONDOX-cal}

\def\l{\left}
\def\r{\right}
\def\gubs{dS_3 \otimes \mathbb{R}}

\DeclareMathOperator{\arctanh}{arctanh}
\DeclareMathOperator{\arccosh}{arccosh}
\DeclareMathOperator{\sech}{sech}

\begin{document}

\count\footins = 1000

\title{Exploring freeze out and flow using exact solutions of conformal hydrodynamics}

\author{Owen Bradley}
\affiliation{           Natural Science Division, Pepperdine University, Malibu, CA 90263, USA }
\author{Christopher Plumberg}
\email{christopher.plumberg@pepperdine.edu}
\affiliation{           Natural Science Division, Pepperdine University, Malibu, CA 90263, USA }

\begin{abstract}
Exact solutions to the equations of hydrodynamics provide valuable benchmark tests for numerical hydrodynamic codes and also provide useful insights into the nature of hydrodynamic flow.  In this paper, we introduce two novel, closely related exact solutions with non-trivial rapidity dependence which are generalizations of the well-known Gubser flow solution to conformal hydrodynamics.  We then use one of our solutions to explore the consequences of choosing between two different criteria for implementing the freeze out process in fluid dynamical simulations of nuclear collisions: freeze out at constant temperature vs. freeze out at constant Knudsen number.  We find that, employing our exact solution, the differences between these freeze out criteria are heavily influenced by the presence of strong collective flow.  Our results highlight the importance of accurately describing the freeze out process in collisions with large flow gradients, particularly in small systems.
\end{abstract}
\maketitle

\section{Introduction}

Relativistic fluid dynamics has proved to be an extremely powerful tool for the description of high-energy nuclear collisions \cite{Gale:2013da, Jeon:2015dfa, Romatschke:2017ejr, Nagle:2018nvi, Adolfsson:2020dhm}.  To date, hydrodynamic models have been used to quantitatively describe a host of experimental observables \cite{Song:2010mg, Bozek:2012qs, Gardim:2012yp, Bozek:2013uha, Niemi:2015qia, Ryu:2015vwa, McDonald:2016vlt, Bernhard:2016tnd, Gardim:2016nrr, Alba:2017hhe, Giacalone:2017dud, Eskola:2017bup, Weller:2017tsr, Schenke:2019ruo, JETSCAPE:2020shq}.  With its ability now to make predictions which are accurate at the percent level \cite{Noronha-Hostler:2015uye, Niemi:2015voa, Adam:2015ptt}, the fluid dynamical paradigm has arguably begun to enter the domain of precision physics.

One of the remaining systematic uncertainties affecting nuclear collision phenomenology involves the treatment of the freeze out process which terminates the fluid dynamical evolution.  The most common approach is to freeze out along a hypersurface of fixed temperature \cite{Heinz:2004qz, Schnedermann:1992ra} or energy density \cite{Petersen:2008dd}, where the latter allows a straightforward generalization to finite chemical potentials which are probed by nuclear collisions \cite{Denicol:2018wdp, Plumberg:2023vkw}.  An alternative (although less widely used) criterion for imposing freeze out is to terminate at constant Knudsen number $\mathrm{Kn}$ \cite{Jeon:2015dfa}.  The Knudsen number represents the competition between mechanisms which drive the system out of equilibrium (e.g., strong flow gradients) and those which relax the system toward local equilibrium (e.g., strong interactions or dissipative effects) and thus provides a measure of how well a system may be described fluid dynamically.  Freeze out at constant Knudsen number can thus be viewed as a choice to decouple the system in terms of the degree to which the fluid dynamical description is justifiable at a given stage in the evolution.

The extent to which the use of different freeze-out criteria influences experimental observables has already received some attention in the context of hydrodynamic models \cite{Schnedermann:1992ra, Eskola:2007zc, Holopainen:2012id, Niemi:2014wta, Ahmad:2016ods, Huovinen:2017hgt}, including systems both with and without event-by-event fluctuations.  However, less work has been done on the differences between these criteria in small systems where large flow gradients drive fluid dynamics to its breaking point \cite{Niemi:2014wta, Heinz:2019dbd, Adolfsson:2020dhm, Summerfield:2021oex, Plumberg:2023fyg}, and relatively little progress has been made in understanding the quantitative implications of these differences for observables which probe specifically the space-time geometry of nuclear collisions, such as Hanbury Brown -- Twiss interferometry \cite{Plumberg:2020jod}.  

In this paper, we begin to address this gap by exploring the extent to which the shape of the freeze out hypersurface is impacted by the choice of freeze out criterion and the presence of collective flow in the system.  In order to build intuition and obtain results which can be straightforwardly interpreted, we have chosen to answer this question in the context of an exact solution to the equations of relativistic fluid dynamics.  The search for such exact solutions has a long history.  Known exact solutions include those obtained by Bjorken assuming longitudinal boost invariance \cite{Bjorken:1982qr} and a later generalization by Gubser \cite{Gubser:2010ze, Gubser:2010ui} to include expansion in the transverse plane as well.  Exact solutions have also been constructed for rotating systems \cite{Hatta:2014gqa, Hatta:2014gga} and systems which possess non-trivial dependence on the space-time rapidity \cite{Csorgo:2018pxh, Shi:2022iyb}, as well as for various generalizations including viscous effects \cite{Marrochio:2013wla}.

In this study, we are interested in exploring the freeze out process using exact solutions which exhibit strong transverse and longitudinal flow simultaneously.  To do this, we first construct two new, exact, Gubser-like solutions which contain non-trivial transverse and longitudinal flow, and then use one of these solutions to compare the geometries of fixed temperature and fixed Knudsen number hypersurfaces.  The first solution (``Solution I") represents a generalization of Gubser's ideal hydrodynamic solution and does not include viscosity, while the second solution (``Solution II") is a generalization of Gubser's viscous solution \cite{Gubser:2010ze, Gubser:2010ui} (and thus allows us to define and study freeze out in terms of the Knudsen number).  Both of our solutions make use of a trick, recently suggested in \cite{Shi:2022iyb} and which we refer to here as a ``temporal shift," to induce a non-trivial rapidity dependence into the collective flow of a known solution.  In this paper we focus our attention on the derivation of the solutions themselves, as well as their qualitative implications for the geometry of the freeze out hypersurface.  We defer a more thorough discussion of the quantitative consequences for nuclear collision phenomenology, such as HBT measurements, to a subsequent study.

We have organized the paper as follows.  In Sec.~\ref{Sec:SolutionI}, we construct Solution I as a generalization of Gubser's ideal flow solution and show how to introduce rapidity dependence by performing a temporal shift.  We also discuss a number of properties of Solution I and consider how its space-time dependence is influenced by different choices of the free parameters it contains.  In Sec.~\ref{Sec:SolutionII}, we apply the same temporal shifting procedure to the viscous version of Gubser's original solution, in order to obtain our Solution II which possesses both viscosity and non-trivial $\eta$-dependence.  Finally, in Sec.~\ref{Sec:Application}, we apply our Solution II to the question of how the presence of both transverse and longitudinal flow influence the shape of the freeze out surface.  Some conclusions are presented in Sec.~\ref{Sec:Conclusions}.

\section{Solution I}
\label{Sec:SolutionI}

Before presenting our derivation of Solution I, we first briefly review the original ideal solution obtained by Gubser \cite{Gubser:2010ze, Gubser:2010ui}.

\subsection{The ideal Gubser solution}
\label{Sec:IdealGubserSolution}

Relativistic fluid dynamics is predicated on the covariant conservation of energy and momentum, expressed by
\begin{align}
\nabla_\mu T^{\mu\nu} = 0 \label{conservations_EoMs}
\end{align}
where $\nabla_\mu$ represents a covariant derivative and $T^{\mu\nu}$ is the energy-momentum tensor.  This yields 4 (3 space, 1 time) equations of motion which dictate the evolution of the system.  For most phenomenological applications, the initial states of nuclear collisions and their subsequent evolution are sufficiently complicated that they can only be solved numerically.  However, in certain highly symmetric scenarios, it may be possible to find exact solutions with at least some phenomenological relevance to nuclear collisions.  The challenge is then to identify a suitable symmetry group which mimics the dynamics of nuclear collisions sufficiently closely.

The original solution obtained by Gubser was constructed in order to respect the symmetry group $SO(3)_q \times SO(1,1) \times \mathbb{Z}_2$, where $SO(1,1)$ represents boost-invariance along the beam axis, $\mathbb{Z}_2$ represents a reflection symmetry under $\eta \to -\eta$, and $SO(3)_q$ denotes a conformal symmetry which includes azimuthal rotations around the beam axis as a subgroup and is designed in such a way as to generate a non-trivial flow profile in the radial direction.  Since the generators of the $SO(3)_q$ and $SO(1,1)$ commute, one thus obtains a solution to the hydrodynamic equations \eqref{conservations_EoMs} which exhibits radial flow while also respecting boost invariance.

The $\gubs$ frame is parameterized by the target space coordinates $\hat{x}^\mu = (\rho, \theta, \phi, \eta)$, while the physical Milne coordinates are $x^\mu = (\tau, r, \phi, \eta)$.\footnote{Here and below, hatted quantities (e.g., $\hat{X}$) are defined in the $\gubs$ frame, while quantities with out hats (e.g., $X$) are defined in physical Milne coordinates.}  In both sets of coordinates, $\phi$ represents the azimuthal angle around the beam axis and $\eta$ represents the longitudinal rapidity, while in Milne coordinates, $\tau$ and $r$ represent the Bjorken proper time and radial coordinate, respectively.  Then the $\gubs$ coordinates $\rho$ and $\theta$ are related to Milne coordinates by
\begin{align}
    \sinh \rho &= \frac{q^2\l( \tau^2 - r^2 \r) - 1}{2 q \tau} \label{rho_definition} \\
    \tan \theta &= \frac{2q r}{1 + q^2\l( \tau^2 - r^2 \r)} \label{theta_definition}
\end{align}
and thus offer an alternative parametrization of the $(\tau,r)$ directions.  In terms of the $\gubs$ coordinates, boost invariance can be expressed as independence of the rapidity $\eta$, while the $SO(3)_q$ symmetry amounts to independence of the coordinates $(\theta,\phi)$.  The $\gubs$ frame has the metric
\begin{align}
\hat{g}_{\mu\nu} = \mathrm{diag}\l( 1, -\cosh^2\rho, -\cosh^2\rho\sin^2\theta, -1 \r) \label{Gubser_metric}
\end{align}
while in Milne coordinates the metric is
\begin{align}
g_{\mu\nu} = \mathrm{diag}\l( 1, -1, -r^2, -\tau^2 \r) \label{Milne_metric}
\end{align}

To obtain Gubser's original solution, one assumes the fluid is stationary in the $\gubs$ frame:
\begin{align}
  \hat{u}_\mu = (1, 0, 0, 0) \label{original_gubs_flow}
\end{align}
By transforming the trivial solution \eqref{original_gubs_flow} back to Milne coordinates, Gubser obtained the solution \cite{Gubser:2010ze, Gubser:2010ui}
\begin{alignat}{3}
    u^\tau &= \tau \frac{\partial \rho}{\partial \tau} &= \frac{1 + q^2\l( \tau^2 + r^2 \r)}{\sqrt{1 + 2q^2\l( \tau^2 + r^2 \r) + q^4\l( \tau^2 - r^2 \r)^2}} \label{idealGubser_utau} \\
    u^r &= -\tau \frac{\partial \rho}{\partial r} &= \frac{2 q^2\ r \tau}{\sqrt{1 + 2q^2\l( \tau^2 + r^2 \r) + q^4\l( \tau^2 - r^2 \r)^2}} \label{idealGubser_ur}
\end{alignat}
with a corresponding energy density $\epsilon$ given by
\begin{align}
    \epsilon &= \frac{\hat{\epsilon}_0}{\tau^{4/3}} \frac{(2q)^{8/3}}{\l[ 1 + 2 q^2(\tau^2+r^2) + q^4(\tau^2 - r^2)^2 \r]^{4/3}} \label{idealGubser_epsilon}
\end{align}
Eqs.~\eqref{idealGubser_utau}-\eqref{idealGubser_epsilon} define Gubser's ideal solution to conformal hydrodynamics.  It contains two free parameters: $q$, which represents an inverse lengthscale characterizing of the flow profile, and $\hat{\epsilon}_0$, which sets the scale for the energy density.

\subsection{Derivation of Solution I}
\label{Sec:Derivation}

We now want to generalize Gubser's solution by relaxing the assumption of boost invariance.  In order to break the boost invariance of the Gubser solution, we may introduce $\rho$ and $\eta$ dependence without violating the $SO(3)_q$ symmetry.  Explicitly, we take the following ansatz for the flow (cf. \cite{Gubser:2010ze, Gubser:2010ui}):
\begin{align}
\hat{u}_\mu = (\cosh \xi(\rho,\eta), 0, 0, \sinh \xi(\rho,\eta)) \label{mod_gubs_flow}
\end{align}
This ansatz reduces to a stationary fluid (and therefore to Gubser's solution) whenever $\xi = 0$.  We emphasize that boost invariance is broken by the inclusion of $\eta$-dependence, but the $SO(3)_q$ symmetry is unaffected.  Note that \eqref{mod_gubs_flow} is properly normalized using the $\gubs$ metric:
\begin{align}
\hat{u}^\mu \hat{u}^\nu \hat{g}_{\mu\nu} = 1
\end{align}

The Gubser solution requires a conformal equation of state.  For simplicity, we assume in this study that $\hat{p} = \hat{\epsilon}/3$.  In this case, the ideal energy-momentum tensor $\hat{T}^{\mu\nu} = \hat{\epsilon} \hat{u}^\mu \hat{u}^\nu + \hat{p}\l( \hat{u}^\mu \hat{u}^\nu - \hat{g}^{\mu\nu} \r)$ has the following non-vanishing components:
\begin{align}
    \hat{T}^{\rho\rho} &= \frac{\hat{\epsilon}}{3}\l( 4\l(\hat{u}^{\rho}\r)^2 - 1 \r) \\
    \hat{T}^{\rho\eta} &= \frac{4\hat{\epsilon}}{3} \hat{u}^\rho \hat{u}^\eta \\
    &= \hat{T}^{\eta\rho} \\
    \hat{T}^{\theta\theta} &= \frac{\hat{\epsilon}}{3}\l( 4\l(\hat{u}^{\theta}\r)^2 + \sech^2\rho \r)
    = \frac{\hat{\epsilon}}{3} \sech^2\rho \\
    \hat{T}^{\phi\phi} &= \frac{\hat{\epsilon}}{3}\l( 4\l(\hat{u}^{\phi}\r)^2 + \sech^2\rho\csc^2\theta \r)
    = \frac{\hat{\epsilon}}{3}\sech^2\rho\csc^2\theta \\
    \hat{T}^{\eta\eta} &= \frac{\hat{\epsilon}}{3} \l(4 \l(\hat{u}^\eta\r)^2 + 1 \r) 
\end{align}
This energy-momentum tensor must be covariantly conserved:
\begin{align}
    \nabla_\mu \hat{T}^{\mu\nu} = \partial_\mu \hat{T}^{\mu\nu} + \Gamma^\mu_{\alpha\mu} \hat{T}^{\alpha\nu} + \Gamma^\nu_{\alpha\mu} \hat{T}^{\alpha\mu} = 0 \label{Tmunu_conservation}
\end{align}
Evaluating the Christoffel symbols for the metric \eqref{Gubser_metric} yields the following non-vanishing components:
\begin{align}
    \Gamma^\rho_{\theta\theta} &= \cosh \rho \sinh \rho, \qquad \Gamma^\rho_{\phi\phi} = \sin^2\theta \cosh \rho \sinh \rho \\
    \Gamma^\theta_{\rho\theta} &= \Gamma^\phi_{\rho\phi} = \tanh \rho, \qquad \Gamma^\theta_{\phi\phi} = -\sin\theta\cos\theta, \qquad \Gamma^\phi_{\theta\phi} = \cot\theta
\end{align}
Then \eqref{conservations_EoMs} implies two independent equations in the $\rho$ and $\eta$ directions:
\begin{align}
    0 &= \partial_\mu \hat{T}^{\mu\rho} + \Gamma^\mu_{\alpha\mu} \hat{T}^{\alpha\rho} + \Gamma^\rho_{\alpha\mu} \hat{T}^{\alpha\mu} \nonumber \\
    &= \partial_\rho \hat{T}^{\rho\rho} + \partial_\eta \hat{T}^{\rho \eta} + 2 \tanh \rho\, \hat{T}^{\rho\rho} + \cosh \rho \sinh \rho \l( \hat{T}^{\theta\theta} + \sin^2\theta \hat{T}^{\phi\phi} \r) \label{EoM1} \\
    0 &= \partial_\mu \hat{T}^{\mu\eta} + \Gamma^\mu_{\alpha\mu} \hat{T}^{\alpha\eta} + \Gamma^\eta_{\alpha\mu} \hat{T}^{\alpha\mu} \nonumber \\
    &= \partial_\rho \hat{T}^{\rho\eta} + \partial_\eta \hat{T}^{\eta\eta} + 2\tanh \rho\, \hat{T}^{\rho\eta} \label{EoM2}
\end{align}
Defining $\hat\epsilon = \exp(4 \mathcal{T})$ and making use of Eq.~\eqref{mod_gubs_flow}, the conservation equations \eqref{EoM1}-\eqref{EoM2} simplify to
\begin{align}
    -\tanh \rho &= \frac{\partial \mathcal{T}}{\partial \rho}
    + \cosh(2 \xi) \l( \tanh \rho + \frac{\partial \xi}{\partial \eta} + 2\frac{\partial \mathcal{T}}{\partial \rho} \r)
    + \sinh(2 \xi)\l( \frac{\partial \xi}{\partial \rho} + 2 \frac{\partial \mathcal{T}}{\partial \eta} \r) \label{most_general_equation_1} \\
    0 &= -\frac{\partial \mathcal{T}}{\partial \eta}
    + \sinh(2 \xi) \l( \tanh \rho + \frac{\partial \xi}{\partial \eta} + 2\frac{\partial \mathcal{T}}{\partial \rho} \r)
    + \cosh(2 \xi) \l(\frac{\partial \xi}{\partial \rho} + 2 \frac{\partial \mathcal{T}}{\partial \eta} \r) \label{most_general_equation_2}
\end{align}
So far no assumptions have been made.  In order to find an exact solution to these equations, we now consider the special case where both the flow rapidity $\xi$ and $\mathcal{T}$ are independent of $\eta$.\footnote{Note that $\eta$ dependence may still be included below by applying a temporal shift, as we describe in Sec.~\ref{Sec:ShiftedSolution}.}  In this case, they can depend only on $\rho$, so that the equations simplify further to
\begin{align}
    0 &= 2 \cosh^2 \xi \tanh \rho + \l( 1 + 2 \cosh\l( 2\xi \r) \r) \frac{d\mathcal{T}}{d\rho} + \sinh\l( 2\xi \r) \frac{d\xi}{d\rho} \\
    0 &= \sinh\l( 2\xi \r) \l( \tanh \rho + 2\frac{d\mathcal{T}}{d\rho} \r) + \cosh\l( 2\xi \r) \frac{d\xi}{d\rho}
\end{align}

The second equation can be solved to find $\mathcal{T}$ in terms of $\rho$ and $\xi(\rho)$; the solution is
\begin{align}
    \mathcal{T}( \rho ) &= c_1 - \frac{1}{2} \ln \cosh \rho - \frac{1}{4}\ln \sinh\l( 2 \xi(\rho) \r),
\end{align}
with $c_1$ an arbitrary constant.  Substituting this solution for $\mathcal{T}$ back into the first equation yields
\begin{align}
    \sinh(2 \xi) \tanh \rho - \l( 2 + \cosh(2\xi) \r) \frac{d\xi}{d\rho} = 0 \label{xi_equation}
\end{align}
This equation is trivially satisfied when $\xi = 0$.  In this case, one would have found the solution for $\mathcal{T}$ to be
\begin{align}
    \mathcal{T}( \rho ) &= \tilde{c}_1 - \frac{2}{3} \ln \cosh \rho,
\end{align}
for some constant $\tilde{c}_1$, implying that
\begin{align}
    \hat{\epsilon} \propto \l( \cosh \rho \r)^{-8/3}, \label{epsilon_ideal_Gubser_scaling}
\end{align}
which is simply Gubser's original solution \cite{Gubser:2010ui}.  The ansatz \eqref{mod_gubs_flow} thus provides a generalization of Gubser flow which reduces to Gubser's ideal solution when $\xi = 0$, as anticipated.

To solve Eq.~\eqref{xi_equation} when $\xi \neq 0$, we recast it into a more manageable form using the changes of variables $\mathcal{x} = \cosh^2\rho$ and $\xi = \frac{1}{2}\ln {\mathcal{y}(\mathcal{x})}$:

\begin{align}
    \frac{d\mathcal{y}}{d\mathcal{x}} = \frac{\mathcal{y}\l( \mathcal{y}^2 - 1 \r)}{\mathcal{x}\l(1 + \mathcal{y}\l( \mathcal{y} + 4 \r) \r)}
\end{align}
It has the solution
\begin{align}
    \mathcal{y}(\mathcal{x}) \equiv {\tilde{\mathcal{y}}\l(c_2\mathcal{x}\r)},
\end{align}
where
\begin{align}
    {\tilde{\mathcal{y}}\l(w\r)} &= 1 - \frac{w}{3} + \frac{2^{4/3} w\l( w - 9 \r)}{a(w)} + 2^{2/3} a(w), \label{solution:y_tilde} \\
    a(w) &= \l[-w\l( 54 + w(2w-27) + 3\sqrt{3 \l(108 - w^2\r)} \r)\r]^{1/3}, \label{solution:a}
\end{align}
and $c_2$ is another arbitrary constant.  Our solution in the $dS_3 \otimes \mathbb{R}$ frame can therefore be written
\begin{align}
    \mathcal{r} &= \cosh^2\rho = \frac{1 + 2q^2\l( \tau^2+r^2 \r) + q^4\l( \tau^2-r^2 \r)^2}{4 q^2 \tau^2} \label{solution:script_r} \\
    \hat{\epsilon}(\rho) &= \frac{2 e^{4c_1}}{\mathcal{r}} \l( \frac{\tilde{\mathcal{y}}(c_2 \mathcal{r})}{\tilde{\mathcal{y}}(c_2 \mathcal{r})^2-1} \r) \label{solution:e_hat} \\
    \xi(\rho) &= \pm\frac{1}{2}\ln {\tilde{\mathcal{y}}(c_2 \mathcal{r})} \label{solution:xi}
\end{align}
 and $c_1$, $c_2 < 0$ are undetermined constants.  Note that the energy density $\hat{\epsilon}$ is conformal; the physical energy density is $\epsilon = \hat{\epsilon}/\tau^4$ in Milne coordinates.  Note also that the rapidity $\xi$ can be chosen positive or negative by the reflection symmetry which is apparent in the above equations.  Here we choose only the positive `branch' of $\xi$ for the sake of definiteness.

 The flow $u^\mu$ \eqref{mod_gubs_flow} can then be mapped to Milne coordinates by the transformation rules \cite{Gubser:2010ui}
 \begin{align}
     u^\tau\l( \tau, r \r) &= \tau \frac{\partial \rho}{\partial \tau} \cosh \xi \label{solution:unshifted_utau} \\
     u^r\l( \tau, r \r) &= -\tau \frac{\partial \rho}{\partial r} \cosh \xi \label{solution:unshifted_ur} \\
     u^\eta\l( \tau, r \r) &= -\frac{1}{\tau} \sinh \xi \label{solution:unshifted_ueta} 
 \end{align}
 We thus arrive at an exact solution for the energy density $\epsilon$ and the flow profile $u^\mu$, given in Eqs.~\eqref{solution:y_tilde}-\eqref{solution:unshifted_ueta}.  Boost invariance has clearly not yet been broken, since the solution is still independent of $\eta$.

 \subsection{Asymptotics}
 \label{Sec:Asymptotics}

 Our exact solution \eqref{solution:y_tilde}-\eqref{solution:unshifted_ueta} is singular in the limit $c_2 \to 0^-$ unless $c_1$ is chosen in an appropriate way (cf.~\eqref{solution:y_tilde} and \eqref{solution:e_hat}).  To avoid this, we define
 \begin{align}
 c_1 = \frac{1}{12} \ln\l( -2 c_2 \hat{\epsilon}_0^3 \r) \label{c1_vs_c2}
 \end{align}
where the parameter $\hat{\epsilon}_0$ is the same one used in Gubser's solution \cite{Gubser:2010ze}.  This combination ensures that the energy density $\hat{\epsilon}$ is regular and reduces to Gubser's original solution as $c_2 \to 0^-$.  One can see this most easily by expanding the solution for $\hat{\epsilon}(\rho)$ in a power series for small $c_2<0$:
\begin{align}
\frac{\hat{\epsilon}(\rho)}{\hat{\epsilon}_0} = \frac{1}{\l( \cosh \rho \r)^{8/3}} - \frac{\alpha}{\l( \cosh\rho \r)^{4/3}} + \frac{3\alpha^2}{4} + O(\alpha^3 \l( \cosh\rho \r)^{4/3}) \label{epsilon_series_expansion}
\end{align}
where we have also defined $\alpha \equiv (-c_2)^{2/3}/(3 \cdot 2^{1/3})$.  The general term in this expansion is clearly of order $\alpha^n \l(\cosh \rho\r)^{4(n-2)/3}$, so that the series reduces to \eqref{epsilon_ideal_Gubser_scaling} when $\alpha^{3/4} \ll \cosh \rho$, i.e., as $c_2 \to 0^-$.

On the other hand, for sufficiently large $\rho$ one finds that the $\tilde{\mathcal{y}}$ decreases inversely with its argument: $\tilde{\mathcal{y}}(w) \propto 1/w$.  Plugging this behavior into \eqref{solution:e_hat}, we find that the conformal energy density scales for large $\rho$ as
\begin{align}    
\hat{\epsilon}(\rho) \sim \mathcal{r}^{-2} \sim e^{-4\rho} \label{SolutionI:epsilon_scaling_large_rho}
\end{align}
One therefore expects a regime in $\rho$ where the leading behavior of $\hat{\epsilon}(\rho)$ transitions from that of Gubser's solution \eqref{epsilon_ideal_Gubser_scaling} to a steeper exponential fall-off in $\rho$ \eqref{SolutionI:epsilon_scaling_large_rho}.  The transition should happen in the neighborhood of some critical value $\rho = \rho_*$ where the sub-leading corrections in \eqref{epsilon_series_expansion} become comparable to the leading term:
\begin{align}    
\frac{1}{\l( \cosh \rho_* \r)^{8/3}} \approx \frac{\alpha}{\l( \cosh\rho_* \r)^{4/3}} \longrightarrow \rho_* \approx \arccosh\l( \alpha^{-3/4} \r) = \arccosh\l( \frac{54^{1/4}}{\sqrt{-c_2}} \r) \label{rho_star}
\end{align}
Interestingly, we note that the definition \eqref{rho_star} implies a constant value for $\xi\l( \rho_* \r) \equiv \xi_* \approx 1.188$, since the $c_2$ dependence cancels in Eq.~\eqref{solution:xi}.  Thus, while the precise ``Gubser time" $\rho_*$ at which the sub-leading corrections become important depends on the value of $c_2$, it occurs at the same flow rapidity $\xi_*$, regardless of $c_2$.

\begin{figure}
    \centering
    \includegraphics[width=0.5\linewidth]{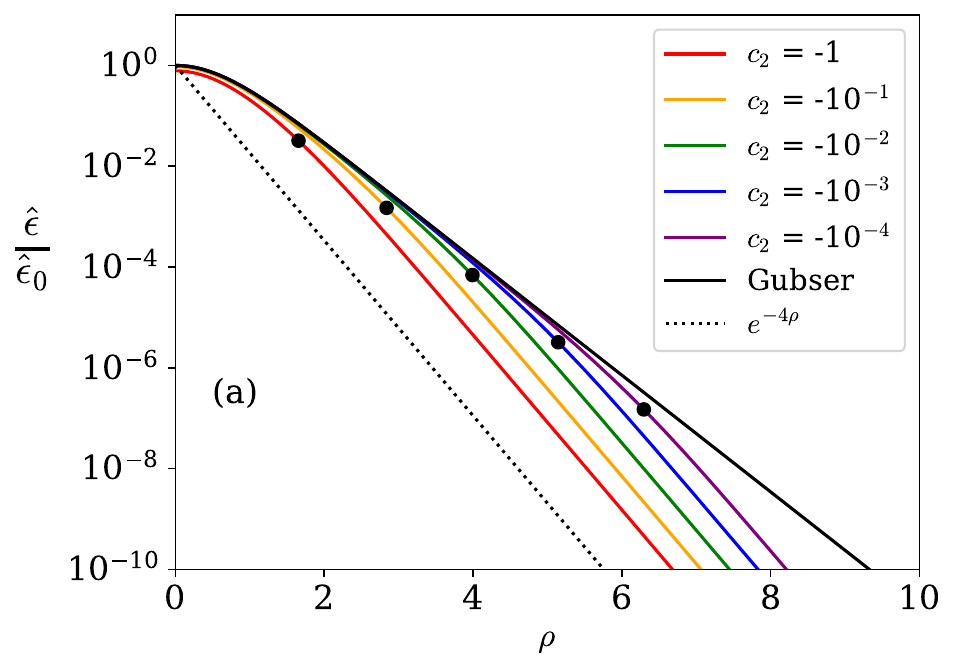}%
    \includegraphics[width=0.45\linewidth]{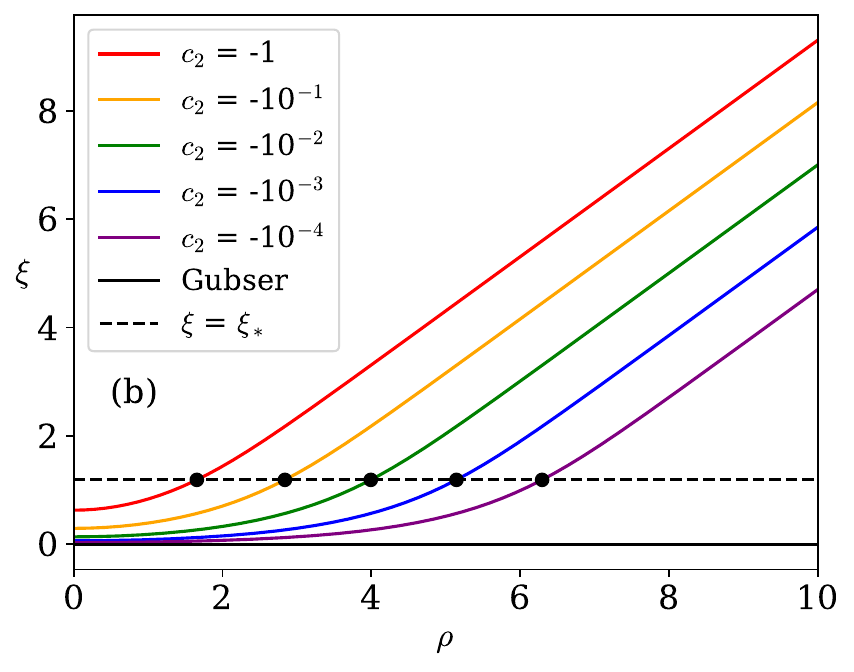}
    \caption{The conformal energy density $\hat{\epsilon}$ (panel (a)) and the flow rapidity $\xi$ (panel (b)) vs. the time-like coordinate $\rho$, for different choices of the free parameter $c_2$.  Gubser's original solution (corresponding to $c_2 = 0$ is given by the thick (solid, black) line in both panels.  The horizontal dashed black line in the righthand panel shows that $\xi_* = \xi\l( \rho_* \r) \approx 1.188$, independent of $c_2$.  See text for discussion.}
    \label{fig:epsilon_and_xi_vs_rho}
\end{figure}

This behavior is illustrated in Fig.~\ref{fig:epsilon_and_xi_vs_rho}, which shows the conformal energy density $\hat{\epsilon}$ (panel (a)) and the flow rapidity $\xi$ (panel (b)) for several choices of $c_2$.  For $\hat{\epsilon}$, one indeed observes that for sufficiently small $\rho$, our solution is close to Gubser's original solution (solid black line), while for sufficiently large $\rho$, the solution instead scales like $e^{-4\rho}$ (black dotted line).  Moreover, the point $\rho_*$ where the transition occurs depends on $c_2$ roughly as implied by \eqref{rho_star}.   For example, for $c_2 = -10^{-3}$, we find using Eq.~\eqref{rho_star} that the critical value $\rho_* \approx 5.14$, in good agreement with Fig.~\ref{fig:epsilon_and_xi_vs_rho}.  Several other values of $\rho_*$ are marked for the various $c_2$ shown.

Fig.~\ref{fig:epsilon_and_xi_vs_rho}(b) shows that $\xi$ is a monotonically increasing function for positive $\rho$, and by symmetry under $\eta \to -\eta$ is monotonically decreasing for negative $\rho$ (cf. \eqref{solution:script_r}).  As $c_2 \to 0^-$, however, the solution again converges smoothly back to Gubser's boost invariant solution.  Note that, since $\xi \neq 0$ whenever $c_2 \neq 0$, $u^\eta$ is always non-zero except when $c_2 = 0$.

Finally, we emphasize that, aside from the reflection symmetry $\eta \to -\eta$, our unshifted solution preserves all of Gubser's original symmetries (including rotational invariance around the beam axis and boost invariance).  In the next subsection, we will show how to modify this solution to break boost invariance as well.

\subsection{The shifted solution}
\label{Sec:ShiftedSolution}

The new solution found above consists of a non-trivial solution for $\epsilon$ and $u^\mu$ in Milne coordinates, which introduces a non-vanishing $u^\eta$ while remaining independent of $\eta$.  In this sense, the solution is analogous to an azimuthally invariant system which nevertheless rotates in the azimuthal direction.  However, our original goal was to construct a solution without boost invariance.  As pointed out in Ref.~\cite{Shi:2022iyb}, additional solutions possessing non-trivial $\eta$ dependence can be generated by simply translating a known solution by a constant amount in the Minkowski temporal coordinate: $t \to t' = t + t_0$.  We call this transformation a ``temporal shift."  This is equivalent to a change of Milne coordinates given by
\begin{align}
    \tau \to \tau'\!\l( \tau, \eta, r \r) &= \sqrt{\tau^2 + 2 t_0 \tau \cosh \eta + t_0^2} \label{tau_prime} \\
    \eta \to \eta'\!\l( \tau, \eta, r \r) &= \arctanh\l(\frac{\tau \sinh \eta}{\tau \cosh \eta + t_0}\r) \label{eta_prime} \\
    r \to r'\!\l( \tau, \eta, r \r) &= r \label{r_prime}
\end{align}
Notice that the transformation on the $r$ coordinate is trivial, since the temporal shift affects only $\tau$ and $\eta$, but we include a shifted $r'$ as well for clarity below.

By definition, the energy density $\epsilon$ transforms as a scalar:\footnote{Here and below, quantities marked with primes ($'$) correspond to the unshifted frame; unprimed quantities are defined in the shifted frame.}
\begin{align}
    \epsilon\l( \tau, r, \eta \r) = \epsilon'\!\l( \tau', r \r) \label{NBI_Gubser_e}
\end{align}
where here and below, $\tau'$ is defined as in Eq.~\eqref{tau_prime}.  Since the flow velocity is a four-vector, it transforms as
\begin{align}
    u^{\tau'} &= \frac{\partial \tau'}{\partial \tau} u^\tau + \frac{\partial \tau'}{\partial \eta} u^\eta + \frac{\partial \tau'}{\partial r} u^r \label{transform_NBI_Gubser_utau} \\
    u^{\eta'} &= \frac{\partial \eta'}{\partial \tau} u^\tau + \frac{\partial \eta'}{\partial \eta} u^\eta + \frac{\partial \eta'}{\partial r} u^r \label{transform_NBI_Gubser_ueta} \\
    u^{r'} &= \frac{\partial r'}{\partial \tau} u^\tau + \frac{\partial r'}{\partial \eta} u^\eta + \frac{\partial r'}{\partial r} u^r \label{transform_NBI_Gubser_ur}
\end{align}
Thanks to the triviality of the transformation in the $r$ direction, the radial component of the flow velocity transforms only by shifting the coordinates in Eq.~\eqref{solution:unshifted_ur} appropriately:
\begin{align}
    u^r(\tau, r, \eta) &= u^{r'}\!\l( \tau', r \r) \label{NBI_Gubser_ur}
\end{align}
The remaining components of the flow velocity are then found by solving \eqref{transform_NBI_Gubser_utau} and \eqref{transform_NBI_Gubser_ueta} for $u^\tau$ and $u^\eta$.  The result is (with $c_1$ set as in Eq.~\eqref{c1_vs_c2}):
\begin{align}
    \mathcal{r}' &= \cosh^2\rho' = \frac{1 + 2q^2\l( {\tau'}^2+r^2 \r) + q^4\l( {\tau'}^2-r^2 \r)^2}{4 q^2 {\tau'}^2} \label{shifted_solution:script_r} \\
    \frac{\hat{\epsilon}(\rho')}{\hat{\epsilon}_0} &= \frac{\l( -16 c_2 \r)^{1/3}}{\mathcal{r}'} \l( \frac{\tilde{\mathcal{y}}(c_2 \mathcal{r}')}{\tilde{\mathcal{y}}(c_2 \mathcal{r}')^2-1} \r) \label{shifted_solution:epsilon_hat}  \\
    \xi(\rho') &= \frac{1}{2}\ln {\tilde{\mathcal{y}}(c_2 \mathcal{r}')} \label{shifted_solution:xi} \\
    u^\tau(\tau,\eta,r) &= \l(\frac{\tau + t_0 \cosh \eta}{{\tau'}}\r) u^{\tau'}\!\l( {\tau'}, r \r) - \l(t_0 \sinh \eta\r) u^{\eta'}\!\l( {\tau'}, r \r) \label{shifted_solution:utau} \\
    u^r (\tau, \eta, r) &= u^{r'} ({\tau'}, r)  \label{shifted_solution:ur} \\
    u^\eta (\tau, \eta, r) &= \l( 1 + \frac{t_0}{\tau} \cosh \eta \r)u^{\eta'}\!\l( {\tau'}, r \r)
    - \frac{t_0}{\tau}  \l(\frac{\sinh \eta}{{\tau'}}\r) u^{\tau'}\!\l( {\tau'}, r \r) \label{shifted_solution:ueta} 
\end{align}
Eqs.~\eqref{shifted_solution:script_r}-\eqref{shifted_solution:ueta}, together with \eqref{tau_prime}, constitute our complete new Solution I which generalizes Gubser flow to include non-boost-invariant longitudinal flow, where the unshifted flow components $u^{\tau'}$, $u^{r'}$, and $u^{\eta'}$ have been defined in \eqref{solution:unshifted_utau}-\eqref{solution:unshifted_ueta}.

\subsection{Properties of Solution I}
\label{Sec:SolutionIProperties}

We now consider how our Solution I obtained above depends on the parameters $c_2$ and $t_0$.  For definiteness, all other parameters are fixed to their values as defined in Refs.~\cite{Gubser:2010ze, Gubser:2010ui}.  Specifically, we take
\[
    q = \l( 4.3 \text{ fm} \r)^{-1} \text{ and } \hat{\epsilon}_0 = 880
\]
which are chosen in order to have a reasonable description of central Au+Au collisions with $\sqrt{s_{NN}} = 200 A$ GeV.

\begin{figure}
    \centering
    \includegraphics[width=\linewidth]{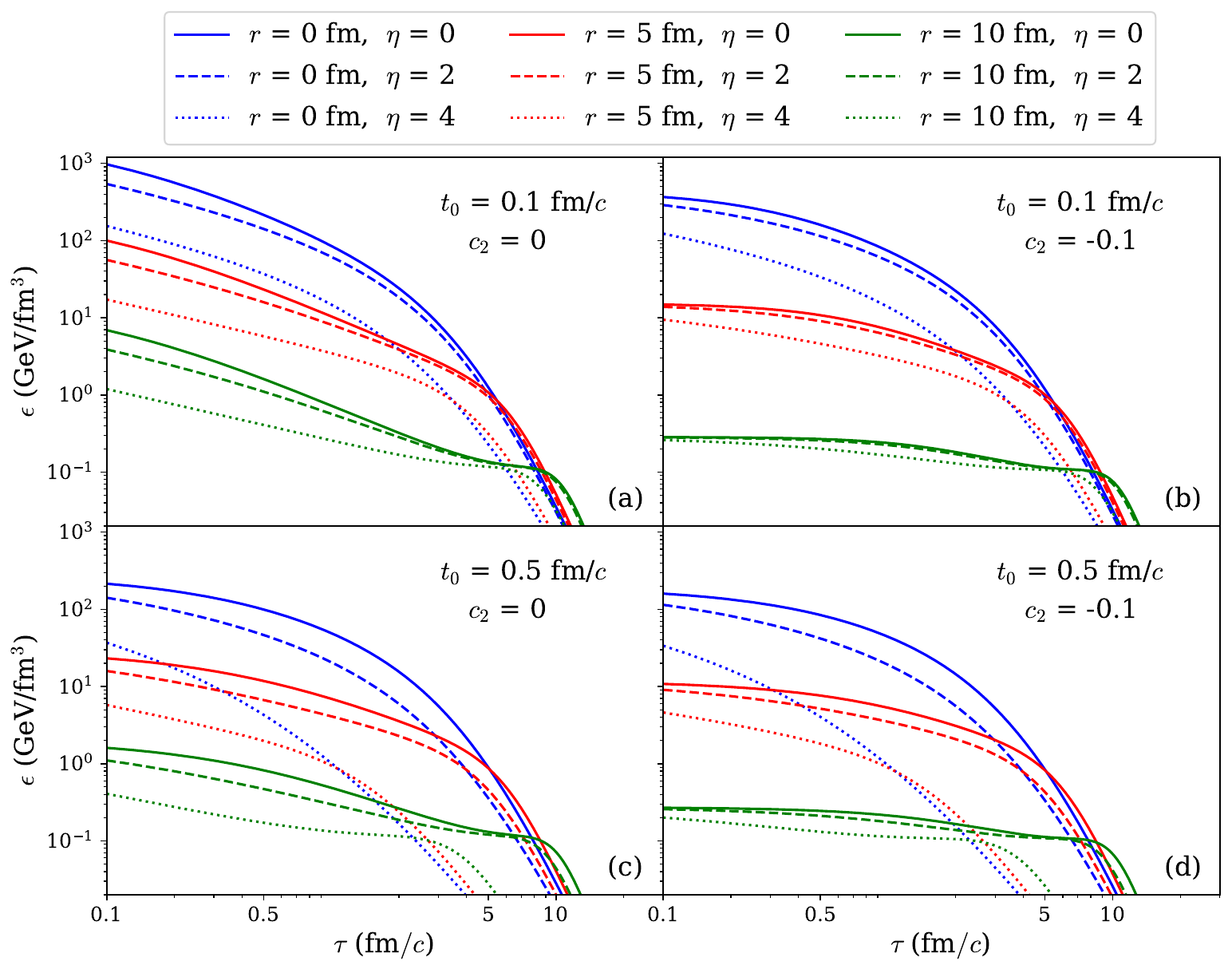}
    \caption{The physical energy density $\epsilon$ as a function of $\tau$, for various radius $r$ and rapidity $\eta$.  Each panel shows a different combination of free parameters $t_0$ and $c_2$:
    $t_0 = 0.1$ fm/$c$ and $c_2 = 0$ (panel (a));
    $t_0 = 0.1$ fm/$c$ and $c_2 = -0.1$ (panel (b));
    $t_0 = 0.5$ fm/$c$ and $c_2 = 0$ (panel (c));
    $t_0 = 0.5$ fm/$c$ and $c_2 = -0.1$ (panel (d)).}
    \label{fig:epsilon_vs_tau}
\end{figure}

We begin by considering how the physical energy density depends on $c_2$ and $t_0$.  This is shown in Figure \ref{fig:epsilon_vs_tau}.  We observe several qualitative behaviors.  First, by comparing panels (a) and (b) with the panels (c) and (d), we see that increasing $t_0$ has the effect of steepening the rapidity dependence of the shifted solution.  This is intuitively plausible, since setting $t_0 = 0$ must lead back to an $\eta$-independent solution.  A similar effect is observed by making $c_2$ non-zero: the rapidity dependence is weakened as the value of $c_2$ is decreased.  Notice further that the effects of modifying $t_0$ or $c_2$ affect $\epsilon$ in different ways: increasing $t_0$ causes $\epsilon$ to fall faster at late $\tau$, while decreasing $c_2$ leads to a faster decrease of $\epsilon$ in the radial direction and only weakly affects the $\tau$ dependence.  We only show curves with $\eta \geq 0$ since the energy density is always symmetric under $\eta \to -\eta$ (cf. \eqref{shifted_solution:epsilon_hat} and \eqref{tau_prime}).

\begin{figure}
    \centering
    \includegraphics[width=\linewidth]{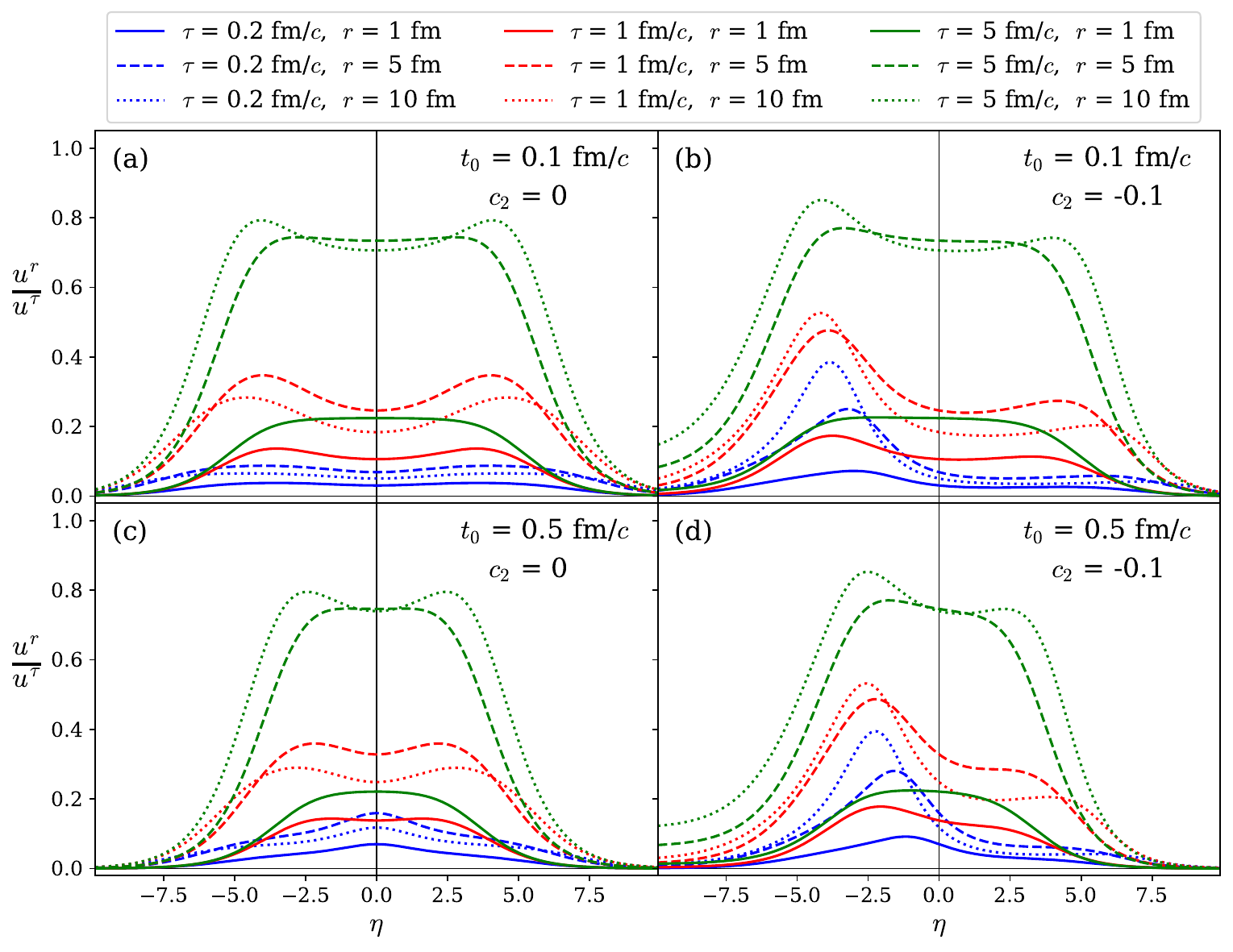}
    \caption{The transverse velocity $u^r/u^\tau$ as a function of $\eta$, for various $\tau$ and $r$.  Each panel shows a different combination of free parameters $t_0$ and $c_2$:
    $t_0 = 0.1$ fm/$c$ and $c_2 = 0$ (panel (a));
    $t_0 = 0.1$ fm/$c$ and $c_2 = -0.1$ (panel (b));
    $t_0 = 0.5$ fm/$c$ and $c_2 = 0$ (panel (c));
    $t_0 = 0.5$ fm/$c$ and $c_2 = -0.1$ (panel (d)).  See text for discussion.}
    \label{fig:ur_utau}
\end{figure}

\begin{figure}
    \centering
    \includegraphics[width=\linewidth]{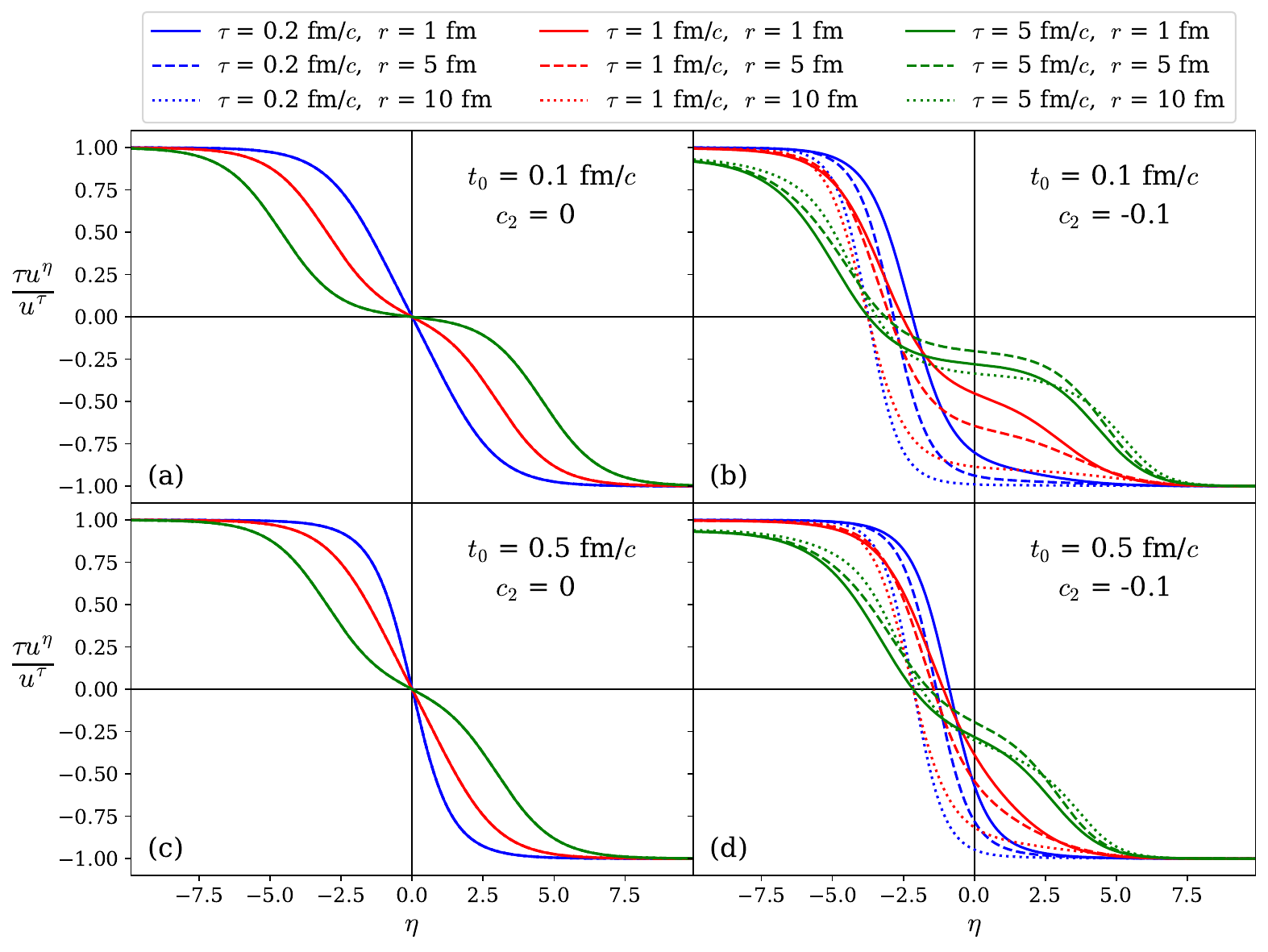}
    \caption{The longitudinal velocity $\tau u^\eta/u^\tau$ as a function of $\eta$, for various $\tau$ and $r$.  Each panel shows a different combination of free parameters $t_0$ and $c_2$:
    $t_0 = 0.1$ fm/$c$ and $c_2 = 0$ (panel (a));
    $t_0 = 0.1$ fm/$c$ and $c_2 = -0.1$ (panel (b));
    $t_0 = 0.5$ fm/$c$ and $c_2 = 0$ (panel (c));
    $t_0 = 0.5$ fm/$c$ and $c_2 = -0.1$ (panel (d)).  See text for discussion.}
    \label{fig:tauueta_utau}
\end{figure}

Our solution's flow profile is shown in Figs.~\ref{fig:ur_utau} and \ref{fig:tauueta_utau}.  In Fig.~\ref{fig:ur_utau}, we show the ratio $u^r/u^\tau$ plotted against rapidity $\eta$, for different combinations of $r$, $\tau$, $t_0$, and $c_2$.  We observe that increasing $t_0$ (cf. panels (a) and (b) with panels (c) and (d)) again reduces the solution's width in $\eta$.  Decreasing $c_2$, however, manifestly breaks the reflection symmetry $\eta \to - \eta$ as anticipated above (cf. panels (a) and (c) with panels (b) and (d)): having chosen the `positive' branch of $\xi$ for our solution in Eq.~\eqref{solution:xi}, we see that the resulting transverse flow is stronger for $\eta < 0$.  This is a consequence of the fact that a positive flow rapidity $\xi$ yields a \textit{negative} rapidity component $u^\eta$, by Eq.~\eqref{solution:unshifted_ueta}.  Choosing $c_2 < 0$ therefore leads to an asymmetric flow in $\eta$, by \eqref{shifted_solution:utau} and \eqref{shifted_solution:ueta}.

In Fig.~\ref{fig:tauueta_utau}, we show the corresponding results for the longitudinal flow velocity, $\tau u^\eta/u^\tau$.  In panels (a) and (c), we see that the only $\eta$ dependence arises from the shift to $t_0 \neq 0$; since $c_2 = 0$, there is no additional rapidity dependence which arises from breaking the reflection symmetry.  In this case $\tau u^\eta/u^\tau$ depends only on $\tau$ and $\eta$, but not on $r$ (cf. Eqs.~\eqref{shifted_solution:utau} and \eqref{shifted_solution:ueta}).  One also observes that increasing $t_0$ steepens the rapidity dependence, as was noted above; setting $t_0 = 0$ would result in $u^\eta = 0$ for all $\tau$ and $r$.  A more nuanced picture of the flow emerges once we let $c_2 < 0$ (shown in Fig.~\ref{fig:tauueta_utau}(b) and (d)).  In this case, the location with the longitudinal flow vanishes shifts to $\eta < 0$, where the size of the shift depending on both $\tau$ and $r$.  One therefore finds that the regions where the rapidity flow ($u^\eta$) is \textit{weakest} are also those where the radial flow ($u^r$) is \textit{strongest}.  This is intuitively what one would expect for an incompressible fluid.  Note, however, that this rough correspondence is not exact: even in the symmetric case ($c_2 = 0$), $u^r/u^\tau$ does not reach a maximum at $\eta=0$ where the longitudinal flow vanishes.  This is due to the complicated nature of Gubser flow and the effects of translating in the temporal direction (cf. panels (a) and (b) in Figs.~\ref{fig:ur_utau} and \ref{fig:tauueta_utau}).

Our solution therefore exhibits some curious properties: the energy density is always symmetric under $\eta \to -\eta$, while the flow profile is asymmetric whenever $c_2 < 0$.  This suggests that scenarios with $c_2<0$ may not bear much phenomenological relevance to the modeling of nuclear collisions, where one would expect to find both flow profiles and energy densities to be asymmetric in $\eta$ \cite{Shi:2022iyb}.  Nonetheless, this solution may also be useful in generating extreme flow profiles which can be used to subject hydrodynamic codes to valuable performance tests.  In the specific case that $c_2 = 0$, however, the temporal shift induces non-trivial $\eta$ dependence without violating the $\mathbb{Z}_2$ symmetry in $\eta$, and thus amounts to a shifted, non-boost-invariant generalization of Gubser's original, boost-invariant solution for ideal hydrodynamics.  In the next section, we show how to apply the same `temporal shift' trick to the viscous version of Gubser's solution, for which the shear viscosity is non-zero.

\section{Solution II}
\label{Sec:SolutionII}

The solution derived above in Sec.~\ref{Sec:SolutionI} was obtained assuming only ideal hydrodynamics.  However, Gubser \cite{Gubser:2010ze} originally derived a version of his boost-invariant solution to the Navier-Stokes equations with a finite specific shear viscosity $\tilde\eta/s \neq 0$.\footnote{Here and below, we denote the shear viscosity by $\tilde\eta$, in order to distinguish it from the space-time rapidity $\eta$.}  Gubser's viscous solution is thus a generalization of our Solution I (with $c_2 = 0$) to a finite shear viscosity.  In this section, we apply a temporal shift to introduce rapidity dependence also into this viscous solution.  This will allow us to define and examine the Knudsen number in Sec.~\ref{Sec:Application}.

Gubser's viscous solution is most conveniently written in terms of the temperature $T_\text{G}$, where the subscript `G' denotes `Gubser.'  $T_\text{G}$ is related to the energy density by $\epsilon \equiv f_* T_\text{G}^4$ (where $f_* \equiv 11$) and is defined by \cite{Gubser:2010ze, Gubser:2010ui}
\begin{align}
    T_\text{G}(\tau, r) &= \frac{\hbar c}{f_*^{1/4} \tau} \hat{T}\!\l(\rho(\tau, r)\r) \label{vGubser_T} \\
    \hat{T}(\rho) &= \frac{\hat{T}_0}{\l(\cosh \rho\r)^{2/3}}\l( 1 + \frac{H_0}{9 \hat{T}_0} \l( \sinh\rho \r)^3 {_2F_1}\l( \frac{3}{2}, \frac{7}{6}, \frac{5}{2}; -\sinh^2\rho \r) \r), \label{vGubser_That}
\end{align}
where $\rho$ is defined by Eq.~\eqref{rho_definition} and $H_0 \propto \eta/s$.  We set $\hat{T}_0 = 5.55$ and $H_0 = 0.33$, corresponding to a choice of $\eta/s = 0.134$ \cite{Gubser:2010ze}.

The flow $u_\text{G}^\mu$ in the viscous case is the same as that obtained in the ideal case \eqref{idealGubser_utau}-\eqref{idealGubser_epsilon}.  One easily checks that $\l(u_\text{G}^\tau\r)^2 - \l(u_\text{G}^r\r)^2 = 1$.  Gubser's complete viscous solution for $T_\text{G}$ and $u_\text{G}^\mu$ is thus given by Eqs.~\eqref{idealGubser_utau}, \eqref{idealGubser_ur}, and \eqref{vGubser_T}.

To obtain the shifted version of Gubser's viscous solution (denoted by the subscript `sG'), we apply the same transformation discussed in Sec.~\ref{Sec:ShiftedSolution}.  Like the energy density $\epsilon$, the temperature $T_\text{G}$ must transform as a scalar.  This implies that Gubser's viscous solution with rapidity dependence is given by
\begin{align}
    T_\text{sG}(\tau, r, \eta) &= T_\text{G}(\tau', r) \label{shifted_viscous_Gubser_T}
\end{align}
In analogy with \eqref{shifted_solution:ur}, the radial component of the shifted flow $u_\text{sG}^\mu$ is
\begin{align}
    u_\text{sG}^r(\tau, r, \eta) &= u_\text{G}^{r'}\l( \tau', r \r) \label{shifted_viscous_Gubser_ur}
\end{align}
The remaining components of the flow velocity are then found by solving \eqref{transform_NBI_Gubser_utau}-\eqref{transform_NBI_Gubser_ueta} for the shifted components as before, and recalling that $u_\text{G}^{\eta'} = 0$ as a result of the boost invariance and reflection symmetry of the Gubser solution.  The result is
\begin{align}
    u_\text{sG}^\tau(\tau, r, \eta) &= u_\text{G}^{\tau'}\l( \tau', r \r) \l(\frac{\tau + t_0 \cosh \eta}{\sqrt{\tau^2 + 2 t_0 \tau \cosh \eta + t_0^2}} \r) \label{shifted_viscous_Gubser_utau} \\
    u_\text{sG}^\eta(\tau, r, \eta) &= -u_\text{G}^{\tau'}\l( \tau', r \r) \l(\frac{t_0 \sinh \eta}{\tau \sqrt{\tau^2 + 2 t_0 \tau \cosh \eta + t_0^2}} \r) \label{shifted_viscous_Gubser_ueta}
\end{align}
Again, one easily confirms that the shifted flow still possesses the correct normalization:
\begin{align}
    \l( u_\text{sG}^\tau \r)^2 - \l( u_\text{sG}^r \r)^2 - \l( \tau u_\text{sG}^\eta \r)^2 = 1
\end{align}
Eqs.~\eqref{shifted_viscous_Gubser_T}-\eqref{shifted_viscous_Gubser_ueta} thus provide a generalization of Gubser's viscous solution which relaxes the assumption of boost invariance by introducing a non-trivial dependence on the space-time rapidity $\eta$.  It coincides with the first solution found above in Secs.~\ref{Sec:SolutionI} only if one takes $c_2 = 0$ and $\tilde\eta/s = 0$.  In Sec.~\ref{Sec:Application}, we will consider several properties of this generalized solution and their implications for the relationship between the freeze out hypersurface and flow in relativistic nuclear collisions.

\section{Flow and Freeze out}
\label{Sec:Application}

Having introduced the two new exact solutions derived above, we turn finally to discuss how we can use them to better understand the role of flow in influencing the freeze out process.  Since we wish to compare freeze out at constant temperature with that at constant Knudsen number, we will focus our attention on Solution II.

In general, the criteria defining freeze out are normally formulated \cite{Jeon:2015dfa} in terms of quantities which depend locally on the hydrodynamic state of the system and which reflect (directly or indirectly) the extent to which the system in a given region may be represented as a strongly coupled plasma of deconfined color charges.  The two criteria we consider here are: (i), freeze out at constant temperature; and (ii), freeze out at constant Knudsen number.  Freeze out at constant temperature has historically been the more widely used criterion, and may be interpreted either as a phenomenological choice of where to `switch' from a partonic to a hadronic description of the system \cite{Ryu:2015vwa, SMASH:2016zqf} or as a way of characterizing the system's passage through the confinement/deconfinement transition \cite{JETSCAPE:2020shq}.  On the other hand, as noted earlier, freeze out at constant Knudsen number characterizes the partonic-hadronic transition in terms of the degree to which a fluid dynamical description is applicable.  Insofar as constant $T$ hypersurfaces closely approximate those of constant $\mathrm{Kn}$ in realistic hydrodynamic simulations \cite{Niemi:2014wta}, the distinction between them is largely academic.  Our goal here, however, is to explore in the context of an analytical solution whether this identification always holds, and if not, under what kinds of conditions it may fail.

The Knudsen number Kn is defined by \cite{Niemi:2014wta, Jeon:2015dfa}
\begin{align}
  \mathrm{Kn} \equiv \frac{\ell_\mathrm{mfp}}{\ell_\mathrm{macro}} \label{Knudsen_definition}
\end{align}
where $\ell_\mathrm{mfp}$ is the mean free path of the fluid and $\ell_\mathrm{macro}$ is an appropriate macroscopic length scale compared to which $\ell_\mathrm{mfp}$ should be sufficiently small in order to justify the applicability of fluid dynamics.

In order to compute the Knudsen number, we must estimate both lengthscales entering Eq.~\eqref{Knudsen_definition}.  In terms of the quantities employed here, $\ell_\mathrm{mfp}$ can be estimated by \cite{Niemi:2014wta,Jeon:2015dfa}
\[\ell_\mathrm{mfp} \sim \frac{5\tilde\eta}{s T}\]
where $T$ is the temperature, and $\tilde\eta/s$ is the ratio of the shear viscosity to entropy density.  This choice of $\ell_\mathrm{mfp}$ corresponds to the value of the shear relaxation time $\tau_\pi$ using the 14-moment approximation of the relativistic Boltzmann equation in the massless limit \cite{Denicol:2012cn, Molnar:2013lta}.  As noted above, since a finite $\ell_\mathrm{mfp} \neq 0$ implies a nonvanishing shear viscosity ratio $\tilde\eta/s \neq 0$, we limit our discussion in this section to the shifted viscous Gubser solution obtained in Sec.~\ref{Sec:SolutionII}.

Many choices are possible for the macroscopic lengthscale $\ell_\mathrm{macro}$ \cite{Niemi:2014wta}; here, we define it in terms of the scalar expansion rate $\theta$,\footnote{Another possible choice is the spatial gradient of the energy density, $\epsilon$, which is expected to be of the same order of magnitude as the definition we use here \cite{Niemi:2014wta}.} which defines a corresponding lengthscale $\ell_\mathrm{macro} = L_\theta$ given by
\begin{align}
    \frac{1}{L_\theta} &\equiv \theta\l( \tau, r, \eta \r) = \partial_\mu u^\mu + \Gamma^\mu_{\mu\alpha} u^\alpha = \frac{1 + q^2\l( r^2 + 5 {\tau'}^2 \r)}{{\tau'} \sqrt{1 + 2 q^2\l( {\tau'}^2 + r^2 \r) + q^4\l( {\tau'}^2 - r^2 \r)^2}} \label{shifted_Gubser_theta}
\end{align}
where $\tau'$ is defined in \eqref{tau_prime}, and the non-vanishing Christoffel symbols in Milne coordinates are
\begin{align}
    \Gamma^\tau_{\eta\eta} = \tau, \qquad   \Gamma^\eta_{\eta\tau} = \Gamma^\eta_{\tau\eta} = \frac{1}{\tau}, \qquad   \Gamma^r_{\phi\phi} = r, \qquad   \Gamma^\phi_{\phi r} =\Gamma^\phi_{r\phi} = \frac{1}{r}
    \label{Christoffel_Milne}
\end{align}
$L_\theta$ thus characterizes the typical lengthscale over which the strength of the system flow changes appreciably. As pointed out above and in \cite{Niemi:2014wta}, these scales should be much larger than the relevant microscopic scale which determines the strength or frequency of interaction, i.e., the mean free path $\ell_\mathrm{mfp}$, in order for fluid dynamics to be applicable.  The Knudsen number may therefore be defined as
\begin{align}
    \mathrm{Kn} &\equiv \mathrm{Kn}_\theta \equiv \frac{\ell_\mathrm{mfp}}{L_\theta} = \frac{5 \tilde\eta \theta}{s T} \label{Knudsen_number_max}
\end{align}
From \eqref{Knudsen_number_max}, it follows that that the Knudsen number is small enough ($\mathrm{Kn} \ll 1$) to warrant the use of fluid dynamics only when either $\tilde\eta/s$ (i.e., out of equilibrium viscous corrections) or $\theta$ (i.e., the flow gradients) is sufficiently small.  In the former case, one may have a violently expanding system which nevertheless is describable using fluid dynamics, as long as the mean free path is sufficiently short and the frequency of interactions sufficiently high.  In the latter case, one may have a relatively large mean free path but nevertheless maintain local equilibrium if the system evolves sufficiently slowly.  Conversely, when either $\tilde\eta/s$ or $\theta$ becomes too large relative to the other, then the interactions become too infrequent to maintain local equilibrium and thus fluid dynamics no longer provides a justifiable description of the system.  It is at this point when freeze out in terms of $\mathrm{Kn}$ should be implemented.

In the remainder of this section, we consider two questions, using our Solution II as a concrete model:
\begin{enumerate}
    \item \textit{How much do the two freeze out criteria (constant temperature, constant Knudsen number) differ in the freeze out contours which they produce?}
    \item \textit{To what extent are these differences influenced by the strength of collective flow and the size of its gradients throughout the system?}
\end{enumerate}

\begin{figure}
    \centering
    \includegraphics[width=\linewidth]{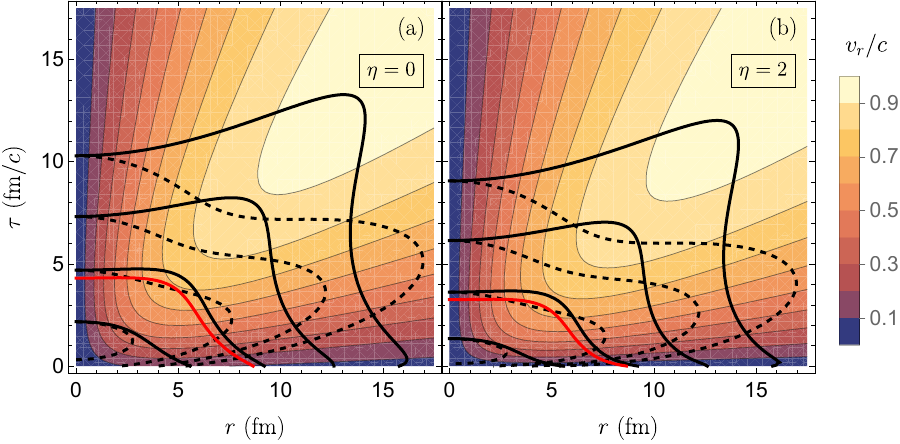}
    \caption{Pairs of freeze out contours using criteria of constant temperature (solid black lines) and constant Knudsen number $\mathrm{Kn}$ (dashed black lines), plotted on top of the transverse flow $v_r = u^r/u^\tau$ in the $\tau$-$r$ plane.  Both panels take $t_0 = 0.5$ fm$/c$ and either $\eta=0$ (panel (a), where $u^\eta = 0$) or $\eta=2$ (panel (b)).  The Knudsen number contours correspond to $\mathrm{Kn} =$ 0.50, 0.75, 1.00, and 1.25 (ordered by increasing $\tau$ at $r=0$), and the temperature contours have been adjusted to match the Knudsen number contours at $r=0$.  A solid red contour at $T = 130$ MeV is included for reference.}
    \label{fig:Kn_plot}
\end{figure}

To address the first question, in Fig.~\ref{fig:Kn_plot} we plot contours of fixed Knudsen number \eqref{Knudsen_number_max} for Solution II (with $t_0 = 0.5 \text{ fm}/c$) as a function of $\tau$ and $r$ (displayed as solid, black lines), for $\eta = 0$ (panel (a)) and $\eta = 2$ (panel (b)).  We also show for reference an isothermal contour of $T = 130$ MeV (solid, red line) corresponding roughly to $\mathrm{Kn} \approx 0.75$, according to the definition $\eqref{Knudsen_number_max}$.  The contours are displayed on top of a density plot of the radial velocity $v_r = u^r/u^\tau$ which indicates where the transverse flow is the strongest.  Four pairs of $T$/$\mathrm{Kn}$ (solid/dashed) contours are shown, where each pair is adjusted to coincide when $r=0$.  The Knudsen contours assume values of $\mathrm{Kn} =$ 0.50, 0.75, 1.00, and 1.25 (corresponding at $r=0$ respectively to $\tau \approx 2.2$, 4.7, 7.3, and 10.3 fm$/c$ and temperatures of $T \approx 209$ MeV, 115 MeV, 69 MeV, and 44 MeV).

One finds that, for $\eta = 0$, $\mathrm{Kn} \lesssim 0.5$ at sufficiently small $r \lesssim 3$ fm and times which are neither too early nor too late ($\tau \lesssim 2$ fm$/c$), and thus that both sets of contours (constant $T$ and constant $\mathrm{Kn}$) demarcate similar \textit{regions} in the $(\tau,r)$-plane where the system may be treated fluid dynamically.  Indeed, the contour of $\mathrm{Kn} = 0.5$ clearly agrees quantitatively with the corresponding contour of constant $T$ at sufficiently small radii $r$ and late times $\tau$.  Similar statements apply to the $\eta=2$ case, except that all freeze out contours are reached at earlier times and thus the regime where fluid dynamics is valid shrinks accordingly.

For larger $\mathrm{Kn}$, Fig.~\ref{fig:Kn_plot} shows explicitly that the \textit{shapes} of constant temperature contours can differ quite significantly, both qualitatively and quantitatively, from contours of constant Knudsen number.  We find that the contours of fixed $T$ acquire a positive concavity at $r=0$ at sufficiently late times and low temperatures.\footnote{In the ideal case, which is a reasonable approximation to the viscous solution at $r=0$, one easily shows that the concavity of the temperature contour changes sign at a critical proper time of $\tau_\mathrm{c} \equiv 1/(c q) = 4.3$ fm$/c$, corresponding to an ideal temperature of $T(\tau_\mathrm{c}) = \hbar c q \hat{T}_0/f_*^{1/4} \approx 140$ MeV.  This seems to agree quite well with where the temperature contours in Fig.~\ref{fig:Kn_plot} become flat at $r=0$.}  This leads to the formation of a `wing-like' structure at large radii which has been noted previously \cite{Heinz:2019dbd} and which has been shown to reflect the violent collective expansion of the system \cite{Plumberg:2020jod, Plumberg:2020jux}.  This is because the interior of the fireball freezes out before the edges do, thanks to the rapid cooling produced by the strong collective flow.

Moreover, one finds that, while the contours corresponding to $\mathrm{Kn} = 0.50$ and 0.75 agree qualitatively with one another, significant discrepancies develop in the contour pairs with $\mathrm{Kn} =$ 1.00 and 1.25 in the region where transverse flow becomes the strongest.  This has a straightforward and intuitive interpretation: since $\mathrm{Kn}$ essentially differs from $T$ only by a constant factor times $\theta$, the differences between constant temperature and constant Knudsen contours must arise from the expansion rate which pushes the constant $\mathrm{Kn}_\theta$ contours to earlier times and smaller radii, compared with constant $T$ contours.  Note that since $u^\eta = 0$ when $\eta = 0$ (by Eq.~\eqref{shifted_solution:ueta}), in this case the discrepancies between the contours are completely determined by $u^r/u^\tau$, as well as the gradients of both transverse and longitudinal flow.

In response to our first question above, then, we conclude that different freeze out criteria such as decoupling at a fixed $T$ vs. a fixed $\mathrm{Kn}$ may indeed lead to dramatically different contours and consequently produce different geometries for the freeze out process.  In the present example, the largest differences occur at larger Knudsen numbers $\mathrm{Kn} \gtrsim 1.00$ and hence smaller temperatures $T \lesssim 100$ MeV.  However, as we have already suggested and now demonstrate, the real origin of these discrepancies is the effect of powerful collective flow with large gradients which drives a rapid cooling of the system and forces an earlier freeze out when it is sufficiently strong.

\begin{figure}
    \centering
    \includegraphics[width=0.33\linewidth]{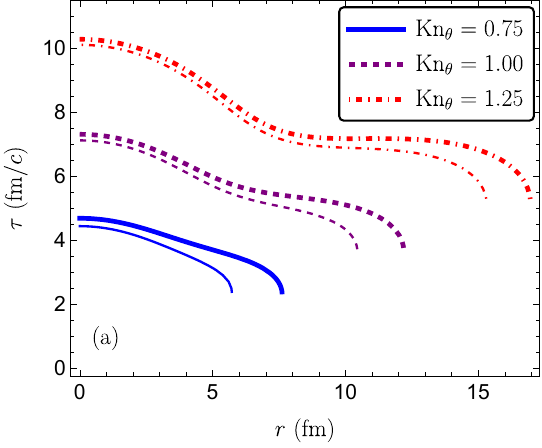}%
    \includegraphics[width=0.33\linewidth]{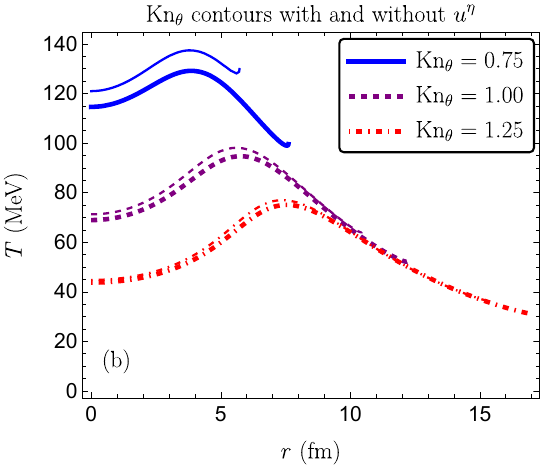}%
    \includegraphics[width=0.33\linewidth]{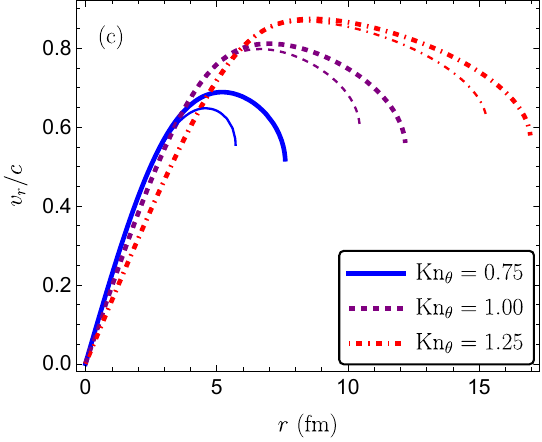}
    \caption{Panel (a) displays thick contours in the $(\tau,r)$-plane for $\mathrm{Kn}_\theta = 0.75$ (solid blue), 1.00 (dashed purple), and 1.25 (dot-dashed red).  For each of these respective contours, we also plot the temperature $T$ (panel (b)) and radial velocity $u^r/u^\tau$ (panel (c)) as a function of radius $r$, with the same styling of lines in each panel.  For each thick contour representing $\mathrm{Kn}_\theta$, we also plot a thin contour of the same color and linestyle to represent $\widetilde{\mathrm{Kn}_\theta}$.}
    \label{fig:quantities_along_Kn_contours_UNPHYSICAL}
\end{figure}

In order to facilitate a direct comparison between the contours of fixed $T$ and fixed $\mathrm{Kn}$ and to show how their differences are influenced by collective flow, in Fig.~\ref{fig:quantities_along_Kn_contours_UNPHYSICAL} we plot the temperature $T$ and radial velocity $v_r$ along the contours of constant $\mathrm{Kn}$ themselves, assuming $t_0 = 0.5 \text{ fm}/c$.  In Fig.~\ref{fig:quantities_along_Kn_contours_UNPHYSICAL}(a), we plot portions of the $\mathrm{Kn}_\theta$ as thick $(\tau,r)$-contours at values of 0.75 (solid blue), 1.00 (dashed purple), and 1.25 (dot-dashed red), extending at late times from $r=0$ to approximately where $d\tau/dr$ along the contour diverges.  Unlike the constant $T$ contours in Fig.~\ref{fig:Kn_plot}, the constant $\mathrm{Kn}$ contours exhibit \textit{negative} concavity at $r=0$: freeze out occurs \textit{last} at the center of the fireball.  In Fig.~\ref{fig:quantities_along_Kn_contours_UNPHYSICAL}(b) and (c), we plot $T$ and $v_r$ along these same contours as functions of $r$ and find that $T$ is decidedly \textit{not} constant along the $\mathrm{Kn}$ contours.  Instead, thanks to the rapid growth of transverse flow with $r$ and its competition with the temperature gradients in the fireball's interior \cite{Schnedermann:1992ra}, one finds that $T$ first \textit{increases} from $r=0$, before reaching a peak and then decreasing toward larger radii.  By inspecting the curves in Fig.~\ref{fig:quantities_along_Kn_contours_UNPHYSICAL}(b), we observe that the relative height of the peak above the central temperature is approximately $13\%$ for $\mathrm{Kn}_\theta = 0.75$, $37\%$ for $\mathrm{Kn}_\theta = 1.00$, and $72\%$ for $\mathrm{Kn}_\theta = 1.25$.  Comparison with the curves in Fig.~\ref{fig:quantities_along_Kn_contours_UNPHYSICAL}(c) then confirms that this relative difference increases with the change in the maximum transverse velocity of the fluid, indicating the strong influence of the latter on the former.  We can therefore answer our second question by confirming that strong collective flow gradients can generate qualitatively significant discrepancies between different freeze out criteria.\footnote{Note that only very slight discrepancies between freeze out criteria were observed in Ref.~\cite{Niemi:2014wta} in the case of p+Pb collisions where the same `wing-like' structure was observed in the constant $T$ contours, in apparent contrast to the results found here.  In this regard, however, we emphasize several key differences between our exact solution and the calculations performed in \cite{Niemi:2014wta}: (i), Gubser's solution (and thus our Solution II) assumes Navier-Stokes hydrodynamics, for which the shear correction $\pi^{\mu\nu}$ is directly proportional to the shear tensor $\sigma^{\mu\nu}$, whereas \cite{Niemi:2014wta} employed Israel-Stewart hydrodynamics \cite{Israel:1979wp}, in which $\pi^{\mu\nu}$ evolves according to a relaxation equation; (ii), \cite{Niemi:2014wta} did not use a constant $\tilde\eta/s$ as assumed here; and (iii), it is unclear whether the Gubser solution's flow gradients are really comparable to those present in the p+Pb simulation shown in \cite{Niemi:2014wta}.  Some combination of these differences may account for the much larger sensitivity to freeze out criterion seen here.}

The tight connection between flow and the freeze out process may be further clarified by considering how the Knudsen contours would be affected by the omission of a portion of the flow.  To see this, we construct an \textit{unphysical} flow which omits the $\eta$ component from $u^\mu$:
\begin{align}
    \tilde{u}^\mu &= \l( \tilde{u}^\tau, \tilde{u}^r, 0, 0 \r) \equiv \l( \sqrt{1 + \l(u^r\r)^2}, u^r, 0, 0 \r) \label{unphysical_flow}
\end{align}
with $u^r$ given by the shifted Gubser solution used in Eq.~\eqref{shifted_Gubser_theta}.  We emphasize that Eq.~\eqref{unphysical_flow} is \textit{not} a solution of the original equations of motion \eqref{conservations_EoMs}, although it remains properly normalized by construction:
\begin{align}
    \tilde{u}_\mu \tilde{u}^\mu = 1
\end{align}
Using $\tilde{u}^\mu$, we compute a corresponding unphysical Knudsen number $\widetilde{\mathrm{Kn}}_\theta$ which may in turn be used to define modified freeze-out contours.  Specifically, we define
\begin{align}
    \widetilde{\mathrm{Kn}}_{\theta} &= \frac{\tilde\eta}{s T} \l( \partial_\mu \tilde{u}^\mu + \Gamma^\mu_{\mu\alpha} \tilde{u}^\alpha \r)
\end{align}
We use the same definition of $T$ for both $\mathrm{Kn}_{\theta}$ and $\widetilde{\mathrm{Kn}}_{\theta}$ as in Eq.~\eqref{shifted_Gubser_theta}.  The corresponding contours are shown as thin lines in Fig.~\ref{fig:quantities_along_Kn_contours_UNPHYSICAL}.  We use three of the same values ($\mathrm{Kn} = 0.75$, 1.00, 1.25) of the Knudsen numbers as before.  By comparing the respective thick and thin line pairs we may then estimate the importance of the component $u^\eta$ for the shape of the freeze out surface.

We first observe the effects of omitting $u^\eta$ on the shape of the $\mathrm{Kn}_\theta$ contours themselves.  By Eq.~\eqref{shifted_solution:ueta}, we see that choosing $t_0 = 0 \text{ fm}/c$ would have resulted in $u^\eta = 0$ and therefore no differences between the $\mathrm{Kn}_\theta$ and $\widetilde{\mathrm{Kn}}_\theta$ contours.  With $t_0 = 0.5 \text{ fm}/c$, however, we see that the unphysical $\widetilde{\mathrm{Kn}}_\theta$ contours shift to smaller $\tau$ and $r$ values relative to the $\mathrm{Kn}_\theta$ contours.  Omitting $u^\eta$ from the shifted viscous Gubser solution of Sec.~\ref{Sec:SolutionII} thus effectively enhances the flow gradients in the system and shrinks the region where fluid dynamics is applicable.  As a result of the artificially shifted contours, the unphysical flow also steepens the temperature and radial flow dependence along the contours.  Looking at the center panel of Fig.~\ref{fig:quantities_along_Kn_contours_UNPHYSICAL}, we find that the discrepancies between the central ($r=0$) and peak temperatures along the contours are slightly enhanced along the shifted $\widetilde{\mathrm{Kn}}_\theta$ contours.  The effects on $v_r/c$ are more pronounced, producing up to a 6\% reduction in the peak radial velocity for the $\mathrm{Kn}_\theta = 0.75$ contour.  Although we do not show it here, the effects of omitting $u^\eta$ are enhanced further as $t_0$ is increased.  We conclude that one effect of setting $t_0>0$ is to `redirect' some amount of the unshifted solution's flow into the rapidity direction, such that artificially suppressing the resulting $\eta$ component of the flow generates spurious gradients which exacerbate the differences between freeze out criteria.  Both transverse and longitudinal flow therefore play essential roles in determining the quantitatively correct shape of the freeze out surface, as well as the variations in the temperature and radial flow along it.

\begin{figure}
    \centering
    \includegraphics[width=0.33\linewidth]{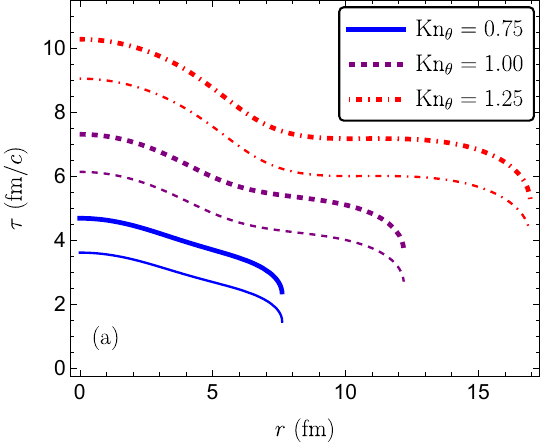}%
    \includegraphics[width=0.33\linewidth]{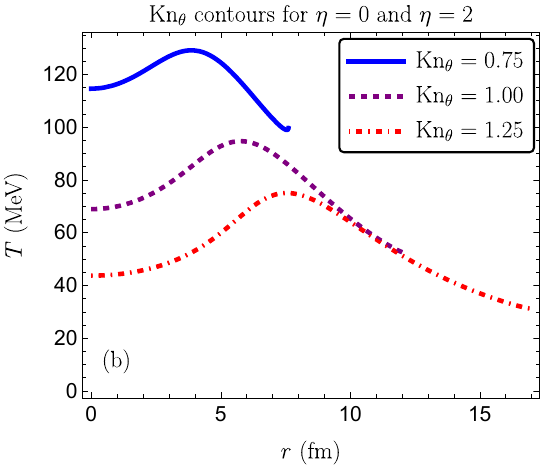}%
    \includegraphics[width=0.33\linewidth]{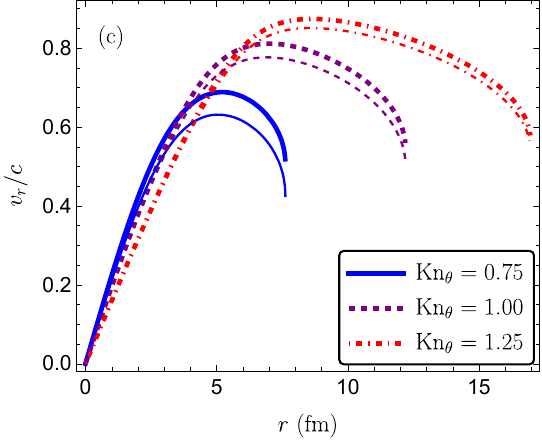}
    \caption{Panel (a) displays contours in the $(\tau,r)$-plane for $\mathrm{Kn}_\theta = 0.75$ (solid blue), 1.00 (dashed purple), and 1.25 (dot-dashed red).  For each of these respective contours, we also plot the temperature $T$ (panel (b)) and radial velocity $u^r/u^\tau$ (panel (c)) as a function of radius $r$, with the same styling of lines in each panel.  Thick contours represent $\mathrm{Kn}_\theta$ at $\eta = 0$, while thin contours represent $\eta = 2$.}
    \label{fig:quantities_along_Kn_contours_RAPIDITY}
\end{figure}

The rapidity dependence itself of the shifted solution can be seen in Fig.~\ref{fig:quantities_along_Kn_contours_RAPIDITY}, where we compare contours of constant $\mathrm{Kn}_\theta$ for $\eta = 0$ (thick lines) with $\eta = 2$ (thin lines).  We emphasize that the solution is independent of $\eta$ (i.e., boost invariant) when $t_0 = 0$.  For $t_0>0$, however, we see in Fig.~\ref{fig:quantities_along_Kn_contours_RAPIDITY}(a) that increasing $\eta$ away from zero shifts freeze out to earlier $\tau$, without dramatically affecting the radial dependence.  The radial velocity (Fig.~\ref{fig:quantities_along_Kn_contours_RAPIDITY}(c)) likewise shifts to smaller values without modifying the radial dependence significantly, producing a roughly 8\% in the maximum $v_r/c$ at $\eta=2$ relative to mid-rapidity when $\mathrm{Kn}_\theta = 0.75$.  In Fig.~\ref{fig:quantities_along_Kn_contours_RAPIDITY}(b), we observe that, since both $T$ and $\mathrm{Kn}$ are Lorentz scalars, their relationship is unaffected by the shift in $t$, and thus the thick and thin lines coincide for all $\mathrm{Kn}_\theta$ contours, regardless of the values of $t_0$ and $\eta$.  Furthermore, increasing the magnitude of $\eta$ further shifts freeze out to still earlier times.  Taken together, these results imply once again that an accurate description of the system's shape and flow are both essential for a quantitatively precise characterization of the freeze out process in nuclear collisions.

Our shifted version of Gubser's original solution thus illustrates the importance of accounting for both transverse and longitudinal flow (and, more precisely, their gradients) on the shape of the freeze out surface.  In particular, the constant temperature contours and constant Knudsen contours imply qualitatively similar geometries for the freeze out hypersurface when the flow is relatively weak.  However, in the presence of strong collective flow, such as one might expect to find in small collision systems \cite{Kalaydzhyan:2015xba, Nagle:2018nvi, Heinz:2019dbd}, which freeze out criterion one adopts is likely to have a larger systematic effect on experimental observables.  The effects could be especially pronounced for observables which are directly sensitive to the geometry of the freeze out hypersurface, such as the Hanbury Brown -- Twiss interferometric radii \cite{Plumberg:2020jod, Plumberg:2020jux}.  We plan to explore the nature of this sensitivity in a subsequent study.

\section{Conclusions}
\label{Sec:Conclusions}

In this work, we have developed two novel exact solutions conformal hydrodynamics.  Both solutions apply a constant shift of the Minkowski time coordinate (a ``temporal shift") to induce a non-trivial rapidity dependence into an already known solution.  In the first case (our ``Solution I"), we have generalized Gubser's solution of ideal relativistic hydrodynamics to allow for flow in the rapidity direction and combined this with a temporal shift to generate a solution with rapidity dependence as well.  In the second case (``Solution II"), we have applied a temporal shift to Gubser's original solution of \textit{viscous} relativistic hydrodynamics, yielding a solution for which we can define out-of-equilibrium quantities like the Knudsen number.  Our new solutions admit a number of interesting applications, including providing useful benchmark tests for 3+1D hydrodynamic codes, as well as allowing analytically tractable insights into the nature of far-from-equilibrium dynamics in nuclear collisions.

As a concrete instance of this latter application, we have used the Solution II constructed here to clarify two related questions: \textit{first}, the amount by which alternative freeze out criteria may differ in the freeze out hypersurfaces which they produce, and \textit{second}, the extent to which these differences are influenced by the strength of collective flow and the size of its gradients throughout the system.

We have answered the first question by considering the qualitative differences which emerge in our solution between contours of fixed temperature $T$ and fixed Knudsen number $\mathrm{Kn}$.  We have found that large discrepancies tend to arise as the system falls progressively away from local equilibrium: in our solution, this tends to occur at temperatures $T \lesssim 100$ MeV.  We anticipate that these differences should lead to systematic effects on the implementation of the freeze out process and the evaluation of, e.g., Cooper-Frye integrals \cite{Cooper:1974mv, Cooper:1974qi} in more realistic numerical simulations.

To address the second question, we directly plotted the local temperature $T$ and radial velocity $v_r$ along contours of constant $\mathrm{Kn}$ using our Solution II.  This allowed us to quantify the typical magnitude of these effects and to explore the role played by collective flow on the discrepancies.  We have also considered how these results are affected by the artificial suppression of $u^\eta$ in the full solution, as a way of assessing their sensitivity to the flow itself.  We found that our solution predicts a strong temperature variation (of up to 72\%) along constant $\mathrm{Kn}$ contours in regions where the flow is sufficiently strong.  Even in regions where the $T$ and $\mathrm{Kn}$ contours were in qualitative agreement, we still observed up to 13\% variation in $T$ (relative to $\mathrm{Kn}$) in those regions where the bulk of particle production occurs \cite{Schnedermann:1992ra}.  This underscores the fact that systematic effects can and do arise on the basis of which freeze out criterion one adopts in hydrodynamic simulations.

Since the discrepancies we observe between different freeze out criteria are somewhat more pronounced than those seen in previous studies (e.g., \cite{Niemi:2014wta, Ahmad:2016ods}), it is important to determine whether the results obtained here can be reproduced in the context of more realistic numerical simulations and to assess what, if any, are the systematic effects on predictions for experimental observables.  Doing so could shed valuable light on the physical mechanisms underlying the robustness of fluid dynamical behavior \cite{Heller:2015dha, Heinz:2019dbd} and help to identify specific observables which are especially well suited to probing the breakdown of hydrodynamics.  In this respect, HBT interferometry is an example of an observable which should be ideal for probing the importance of these effects for the geometry of the freeze out hypersurface, and we plan to explore its sensitivity to the freeze out criterion in a subsequent study.

\acknowledgements

The authors sincerely thank Travis Dore, Ulrich Heinz, and Jorge Noronha for several useful comments and suggestions for improvements to the manuscript.

\bibliography{inspire}

\end{document}